\documentclass[10pt]{article}
\usepackage{amssymb}
\usepackage{axodraw}
\usepackage{rotate}
\usepackage{epsfig}
\usepackage{graphicx}
\usepackage{floatflt}

\def\bqa{\begin{eqnarray}}
\def\eqa{\end{eqnarray}}
\def\bq{\begin{equation}}
\def\eq{\end{equation}}
\def\dd{{\mathrm d}}

\def\mw {M_{\sss W}}

\def\hspr{\hat{s}'}

\def\msbar{\overline{ \mathrm{MS}}}
\def\phAV{\phantom{ AVALUE =}}

\newcommand{\sss}[1]{\scriptscriptstyle{#1}}
\def\ds{\displaystyle}

\begin{document}
\setcounter{page}{0}
\thispagestyle{empty}

$\,$
\vspace*{-1cm}
\vspace*{\fill}
\begin{center}

{\LARGE\bf  Standard {\tt SANC} Modules}
\vspace*{1.5cm}

{\bf A.~Andonov, A.~Arbuzov$^*$, D.~Bardin, S.~Bondarenko$^{*}$, \\[1mm]
\bf P.~Christova, L.~Kalinovskaya, V.~Kolesnikov and R.~Sadykov}

\vspace*{13mm}

{\normalsize
{\it Dzhelepov Laboratory for Nuclear Problems, JINR,      \\
$^{*}$ Bogoliubov Laboratory of  Theoretical Physics, JINR,\\ 
       ul. Joliot-Curie 6, RU-141980 Dubna, Russia,       
}}
\vspace*{13mm}

\end{center}

\begin{abstract}
\noindent
In this note we summarize the status of the standard {\tt SANC} modules
(in the EW and QCD sectors of the Neutral Current branch  --- version {\tt 1.20} and 
    the Charged Current branch --- version {\tt 1.20}).
All versions of the codes are accessible from the {\tt SANC} project servers at Dubna 
{\it http://sanc.jinr.ru} and CERN 
{\it http://pcphsanc.cern.ch}.
\end{abstract}

\vspace*{6cm}

\bigskip
\footnoterule
\noindent
{\footnotesize \noindent
E-mail: sanc@jinr.ru}
\clearpage
\tableofcontents
\listoffigures
\listoftables    
\clearpage

\section{Introduction\label{introduction}}
 The status of the theoretical description of the Drell--Yan--like processes~\cite{Drell:1970wh} 
was widely overviewed in the resent papers~\cite{Baur:2007ub,CarloniCalame:2006bg} 
where the necessity of further in-depth study of them both from experimental and theoretical
sides was emphasized.

 Single W and Z boson production in Drell--Yan-like (DY)  processes have clean
signature and large cross sections.
Their precision studies will be used for determination of the parton density functions,
luminosity, $\mw$, $\sin^2\theta^{eff}$, $\Gamma_{{\sss W},{\sss Z}}$.
Required precision tag is below 1\%, see e.g.~\cite{Dittmar:1997md,Frixione:2004us}.

Theoretical calculations of the DY processes for high energy hadronic colliders were performed at the level 
of one-loop QED and electroweak (EW) radiative corrections (RC) by several groups, see 
papers~\cite{Wackeroth:1996hz,Baur:1998kt,Dittmaier:2001ay,Baur:2002fn,Baur:2004ig,CarloniCalame:2006zq} 
and references therein. QCD corrections are known up to the next-to-next-to-leading order~\cite{Melnikov:2006kv}.

The first attempt to combine EW and QCD corrections was done within the working group 
``The neutral current Drell-Yan process in the high invariant mass region'' of the the 2007 Les Houches 
Workshop~\cite{Buttar:2008jx}. 
Existing codes have to meet the requested experimental precision.
QED, Weak and QCD one-loop level corrections should be calculated taking into account by oneself, 
their interplay and all necessary higher order effects.

For users attention we offer two types of {\tt SANC} outputs
\footnote{We do not describe {\tt SANC} system in this note, referring the reader to published 
papers~\cite{Andonov:2004hi},~\cite{Bardin:2005dp}.}:
 
\begin{itemize}
\item{} stand-alone packages for calculation of the EW and QCD NLO RC at the parton level together with the environment 
in which it is run, i.e. the Standard {\tt SANC} FORM (and/or FORTRAN) Modules (SSFM) (see section \ref{sect:tree}); 
\item{} MC event generators, for production of event distributions at the hadronic 
level, based on the {\tt FOAM} algorithm \cite{Jadach:2005ex}. 
\end{itemize}

 In this note we present the first positive experience of the 
using SSFM's for DY-like processes for neutral current (NC), 
\bqa
q \bar{q} \to \ell^+\ell^-
\label{partlevelCC}
\eqa
with $\ell=\mu,e$, (the NC DY generator was presented first in talk~\cite{Sadykov:2007mc})
and charged current (CC) processes (in --- \cite{Sadykov:2008mc})
\bqa
q \bar{q^{'}} \to \ell \nu_\ell\,,
\label{partlevelNC}
\eqa
as stand-alone codes, see also~\cite{Kolesnikov:2008acat} and~\cite{Arbusov:2008acat}.

 The packages for calculations at the partonic level as well as the {\tt FOAM}-based generators 
are accessible from {\tt SANC} project homepages \cite{SANCsite:2008}.

 Moreover, we create tools for checking the implementation of these modules --- the integrators 
of the processes, based on the Vegas algorithm \cite{Lepage:1977sw}--\cite{Lepage:1980dq}, but 
presently only for an internal use. 
We used these integrators in the business of the tuned comparison within the several 
international workshops for NC and CC DY processes,
see Les Houches Workshop Proceedings  \cite{Buttar:2006zd}, \cite{Buttar:2008jx} 
and Tevatron for LHC Report \cite{Gerber:2007xk}.

 In this note we demonstrate the first example of applying EW SSFM into MC generator 
{\tt WINHAC}~\cite{Bardin:2008fn}.
We checked the work of these modules by means of comparison of the results obtained by {\tt SANC} integrator
DY\_CC\_VEGAS with those produced by the development version {\tt WINHAC v1.30}
(see Section \ref{SANCandWINHAG}).

 The {\tt SANC} project was presented at several ATLAS MC and SM working group meetings,
e.g.~\cite{Bardin:2006at,Sadykov:2006mc,Kolesnikov:2006mc,Bardin:2007sm,Bardin:2008sm}.
From the project page one can download {\tt SANC} client software, that allow users to interactively test the
workability of the all offered processes, i.e. user may create SSFM's by oneself (the  documentation
for {\tt SANC} client also can be found at the project webpage).
However, user may download SSFM from the {\tt SANC} homepages directly.

\clearpage  

\section{Basics of working with {\tt SANC}\label{DYbranches}}

\subsection{Working with {\tt SANC} client \label{sancclient}}
\begin{figure}[!th]
\includegraphics[width=16cm]{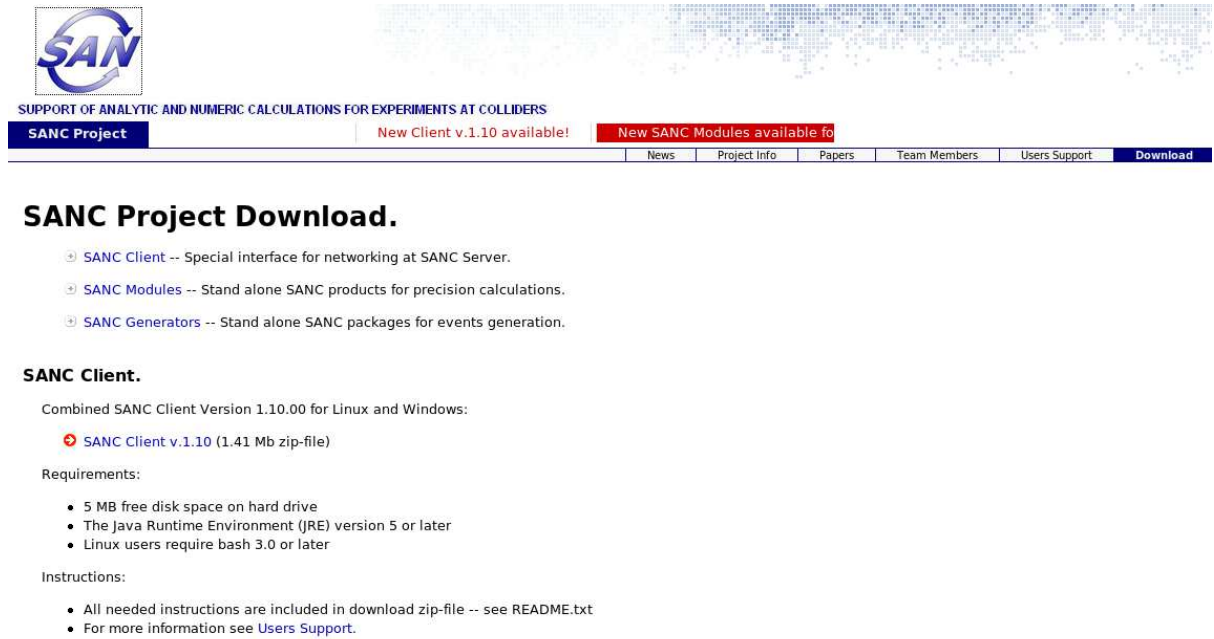}
\caption{{\tt SANC} Project Download}
\label{SANCPD}
\end{figure}

To work with {\tt SANC}, one must install the {\tt SANC} client.
The {\tt SANC} client can be downloaded from the {\tt SANC} project homepage
\cite{SANCsite:2008}.
On the homepage select {\bf Download}, see Fig.\ref{SANCPD}, then download the client,
unzip it and follow the instructions in the README file.\footnote{
To install and run {\tt SANC} client one should
have the Java Runtime Environment (JRE) at least version
5.0 Update 5 installed, see section {\bf Minimum System Requirements}
of the {\bf Download} page at the {\tt SANC} project homepage.}

\subsection{{\tt SANC} tree for Drell--Yan processes \label{sect:tree}}

\begin{figure}[!h]
\begin{center}
\begin{tabular}{cc}
 \includegraphics[width=0.4\textwidth]{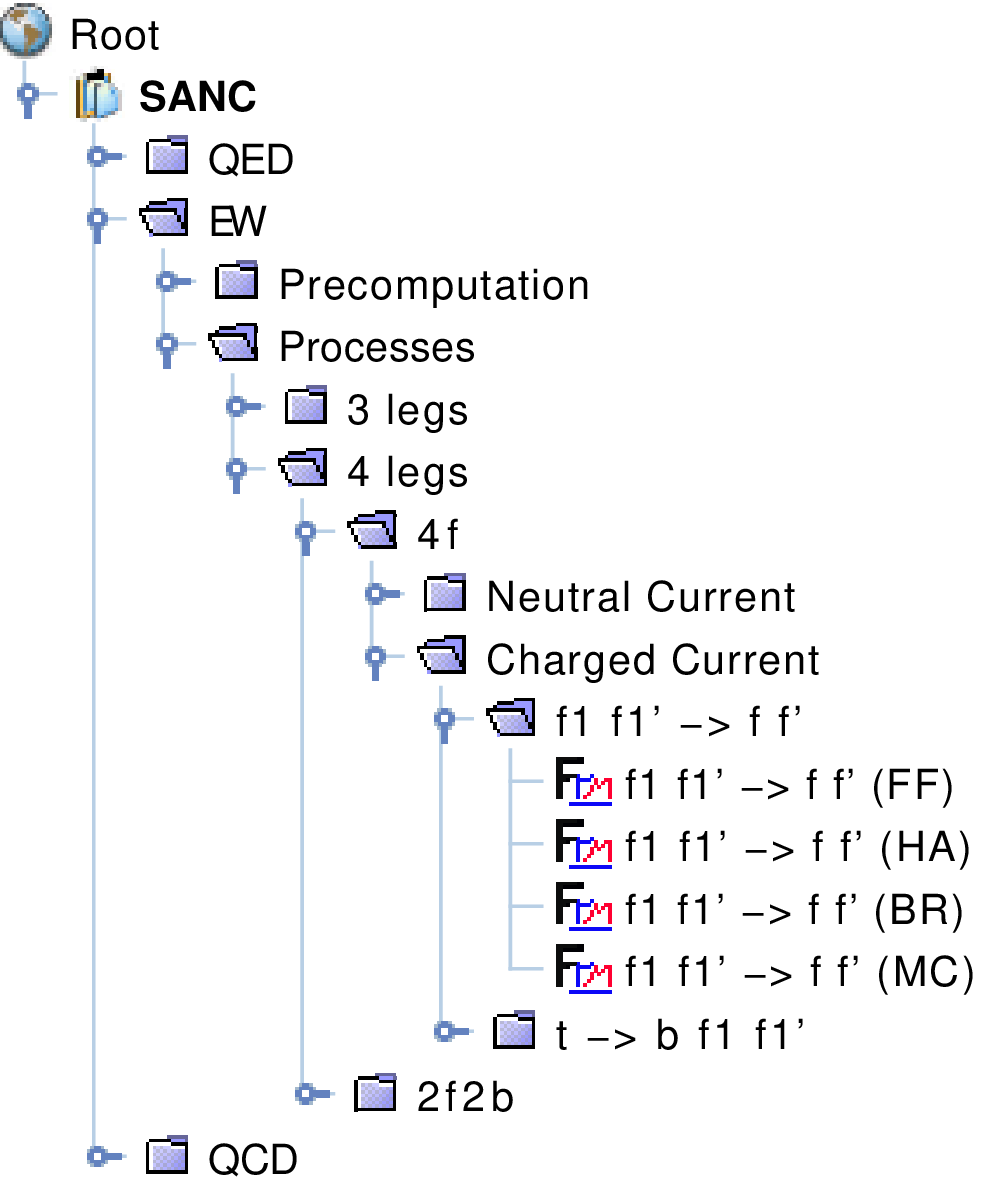}
&\includegraphics[width=0.4\textwidth]{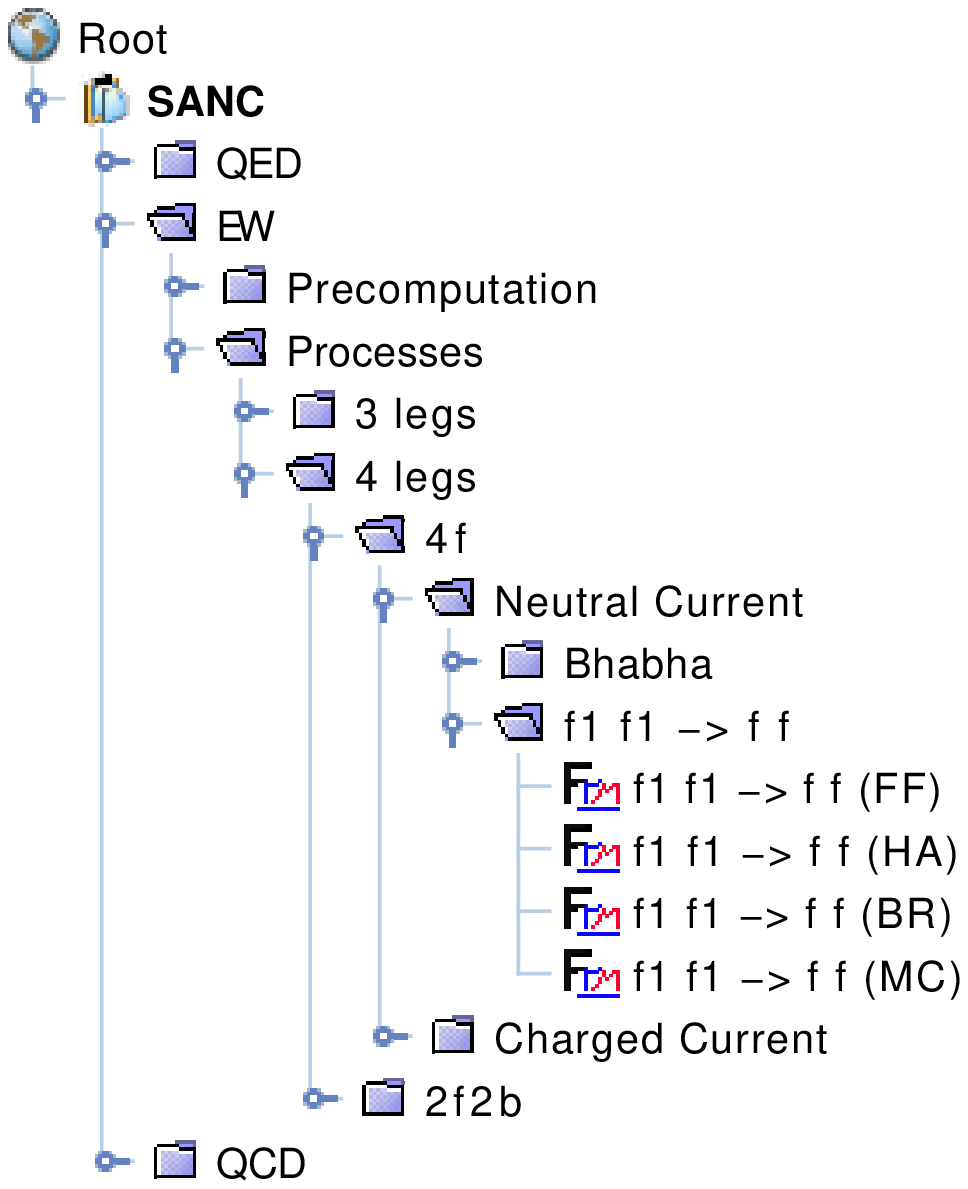}
\end{tabular}
\end{center}
\vspace*{2mm}
\caption{{\tt SANC} tree for the Drell--Yan processes: Charged and Neutral Current case}
\label{sanc:tr} 
\end{figure}

 The desired stage of development of the {\tt SANC} tree for any process
is the tree with four FORM modules at the tip of the processes branches.
In Fig.~\ref{sanc:tr} we show the location of the $2f \to 2f$ CC and NC sub-processes at the {\tt SANC} tree. 

 Moving along the menu sequence 
{\bf SANC $\to$ EW $\to$ Processes $\to$ 4 legs $\to$ 4f $\to$ Neutral Current 
$\to$f1~f1~$\to$~f~f~}({\bf FF,HA,BR,MC}) 
the user can see a list of the \underline{Standard SANC FORM Modules}, the scalar Form Factors ({\bf{FF}});
then at the second module, Helicity Amplitudes ({\bf{HA}}); then at the third module, 
the integrated bremsstrahlung ({\bf{BR}}); and finally at the fourth module, 
the fully differential bremsstrahlung ({\bf{MC}}). 

 The same way exists for the branch of {\bf Charged Current $\to$f1~f1'~$\to$~f~f'~} ({\bf FF,HA,BR,MC})
standard {\tt SANC} modules, see~Fig.~\ref{sanc:tr}. 

 The FORM modules compute online the FF, HA, BR and MC contributions of the corresponding
partonic sub-processes.

 Each of these modules in turn can be opened, compiled and run as explained in the UserGuide\_v1.10 
(see project homepage).

 For more details of running the existing accessible menu, one can refer to the description of the {\tt SANC} 
system in Ref.~\cite{Andonov:2004hi}.

 Let us consider the types of modules. Each FORM module produces an output,
which is in turn an input for creation of the corresponding FORTRAN code by the software {\tt s2n}.

$\bullet$ {\bf FF} and {\bf HA}

In Ref.~\cite{Andonov:2004hi} we presented the Covariant (CA) and Helicity Amplitudes (HA) for
$f_1{\bar f}_1 f{\bar f}\to 0$ NC and CC processes, with all 4-momenta being incoming for
any of its cross channels $s,t$ or $u$.
The expressions for the CA and HA
(see Eq.(30) and (33) of the last reference) of these processes are written in terms of scalar FF's.

$\bullet$ {\bf BR} and {\bf MC}

The {\bf BR} module computes the soft and inclusive hard real photon emission: 
\bqa
\bar q + q \to \ell + {\bar \ell} + \gamma.
\eqa
We do not discuss the soft photon contribution here, referring the reader to 
the system itself. As far as hard photons are concerned, we realized two 
possibilities of the integration over their phase space:
the semi-analytical one (module {\bf BR}) and the one by means of a Monte Carlo integrator 
or generator (module {\bf MC}).

The {\bf MC} module provides fully differential hard brem\-sstrahlung contribution to the 
partonic cross section. 
The contribution is given in a form suitable for further numerical integration or
simulation of events in a Monte Carlo generator.

\clearpage

\subsection{Types of {\tt SANC} Outputs}
As we mentioned in section \ref{introduction}, {\tt SANC} produces two types of outputs: SSFM and 
MC generators at the one-loop level.

\begin{figure}[!h]
\label{ssfm}
\includegraphics[width=16cm]{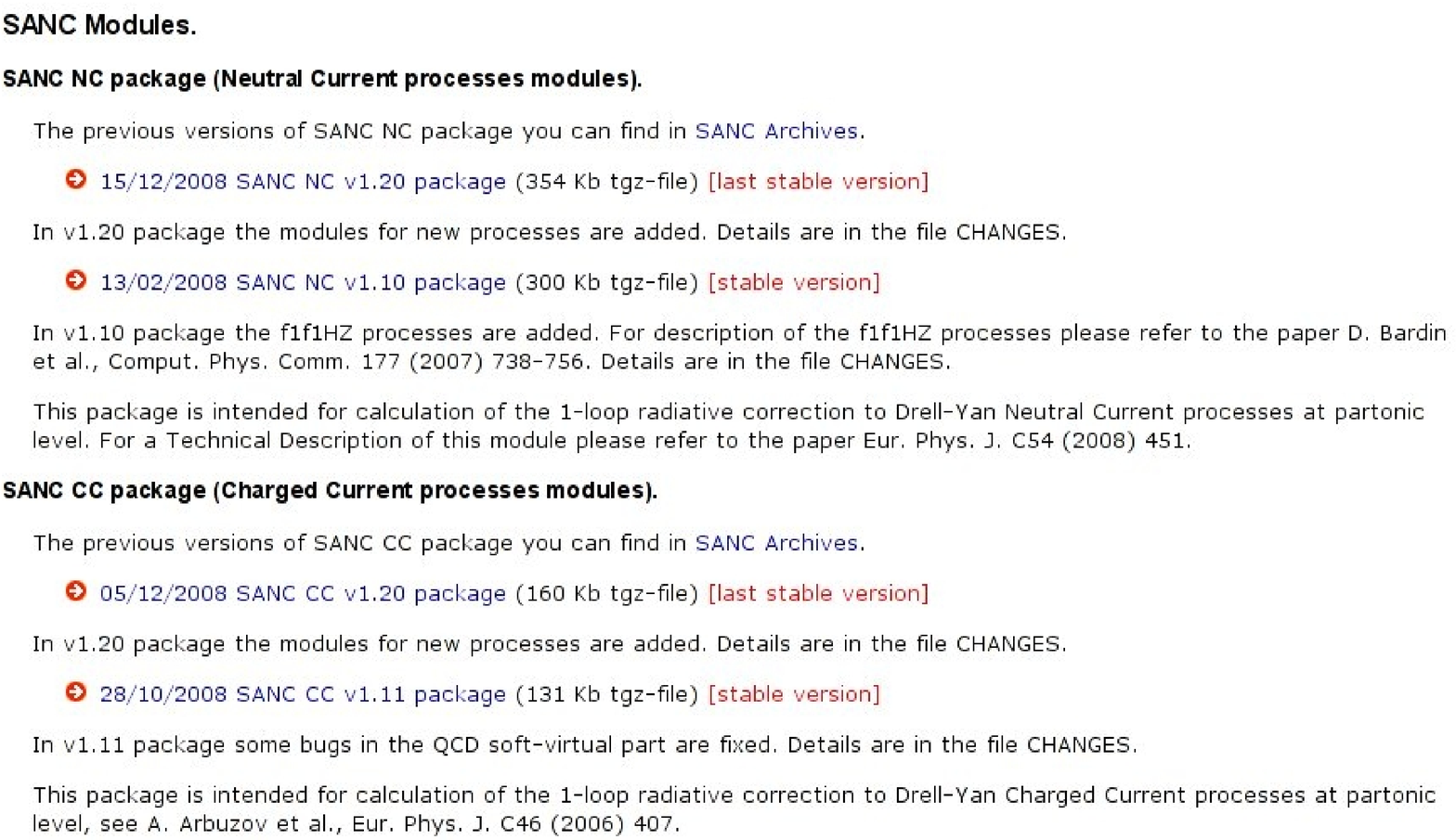}
\caption{Available versions of the SSFM}
\label{SANCmodules}
\end{figure}

$\bullet${~\underline{Standard {\tt SANC} FORTRAN Modules}}, i.e.
the list of packages at the parton level for download is, (see Fig.\ref{SANCmodules})
\begin{itemize} 
\item[-]{{ SANC NC v1.20:}
{SSFMs for the DY NC processes: $(uu,dd)\to (\mu\mu,ee)$}~\cite{Arbuzov:2007db},
{SSFMs for the processes: $ee (uu,dd)\to HZ$}~\cite{Bardin:2005dp}
and NC gluon-induced processes~\cite{Arbuzov:2008in};
     }
\item[-]{{SANC CC v1.20:}
{SSFM for the DY CC processes: $(uu,dd)\to (\mu\nu_{\mu},e\nu_e)$}~\cite{Arbuzov:2005dd} and CC gluon-induced 
processes~\cite{Arbuzov:2008in}.
     }
\end{itemize}

One has to emphasize that the CC v1.10 contains first {\tt SANC} modules for the calculation of NLO QCD 
corrections~\cite{Andonov:2007zz}. Some results about interplay of QCD/EW correction were reported in
talks to ATLAS MC group~\cite{Sadykov:2008mc},\cite{Sadykov:2006mc},\cite{Kolesnikov:2006mc}.

\clearpage  

$\bullet${~\underline{Standalone MC generators:}}

\vspace*{3mm}

\begin{figure}[!h]
\includegraphics[width=16cm,height=10cm]{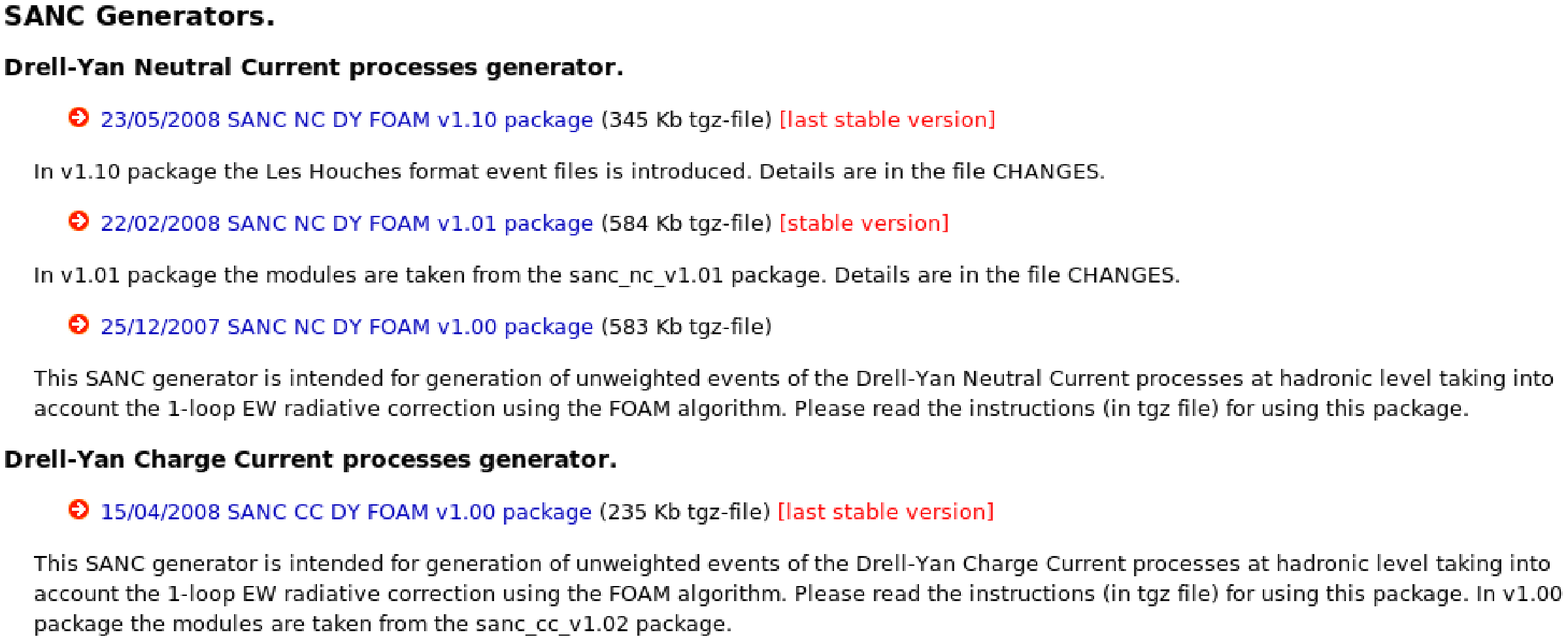}
\vspace*{-10mm}
\caption{{\tt SANC} MC Generators} 
\label{SANCgen}
\end{figure}

The sanc\_$**$\_foam\_v1.** packages are intended for generation of unweighted
events of the DY processes in NC and CC sector at the hadronic
level taking into account the one-loop EW radiative correction
based on {\tt FOAM} algorithm \cite{Jadach:2005ex}, (see Fig.~\ref{SANCgen}).
These generators use the standard {\tt SANC} Fortran modules for calculation of NLO EW corrections
as well as {\tt LoopTools-2.1}~\cite{homepagelooptools} and {\tt SancLib-v1-00} libraries for evaluation of scalar and
tensor one-loop integrals.  Also you need {\tt ROOT} package to be installed at your computer
to use these generators.

In the present version of packages we include the possibility to write the output in data files containing the event
information in the standard Les Houches Accord format \cite{Alwall:2006yp} in order to organize
the transfer of information between {\tt SANC} generators and general purpose programs such as
{\tt PYTHIA} \cite{Sjostrand:2007gs} and {\tt HERWIG} \cite{Bahr:2008pv}. 

We advice to read \underline{INSTALL} for installation instructions, \underline{UserGuide.txt} for a 
Technical Description and \underline{readme\_foam} for {\tt FOAM} using.

\begin{figure}[!th]
\includegraphics[width=8cm,height=5cm]{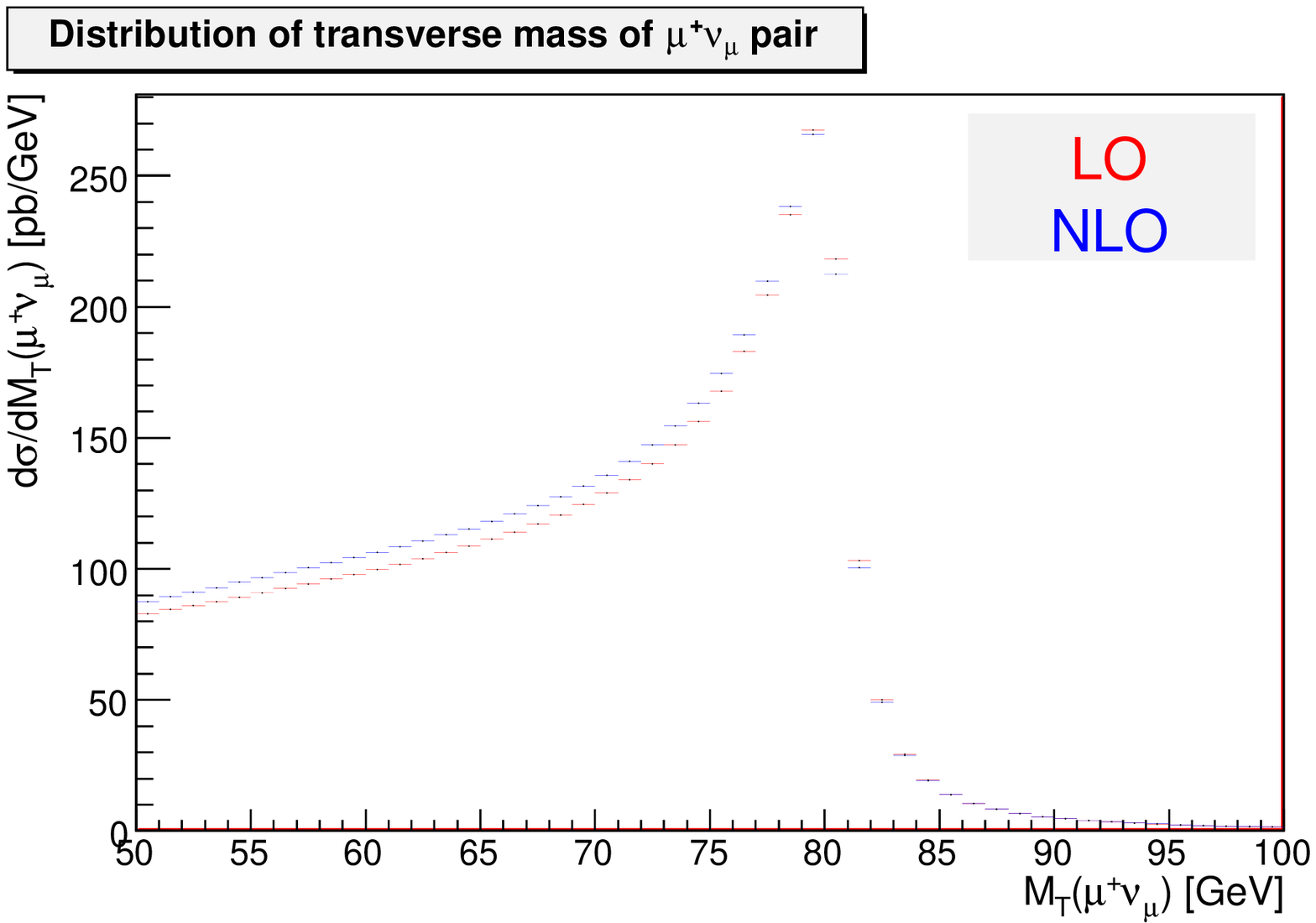}
\includegraphics[width=8cm,height=5cm]{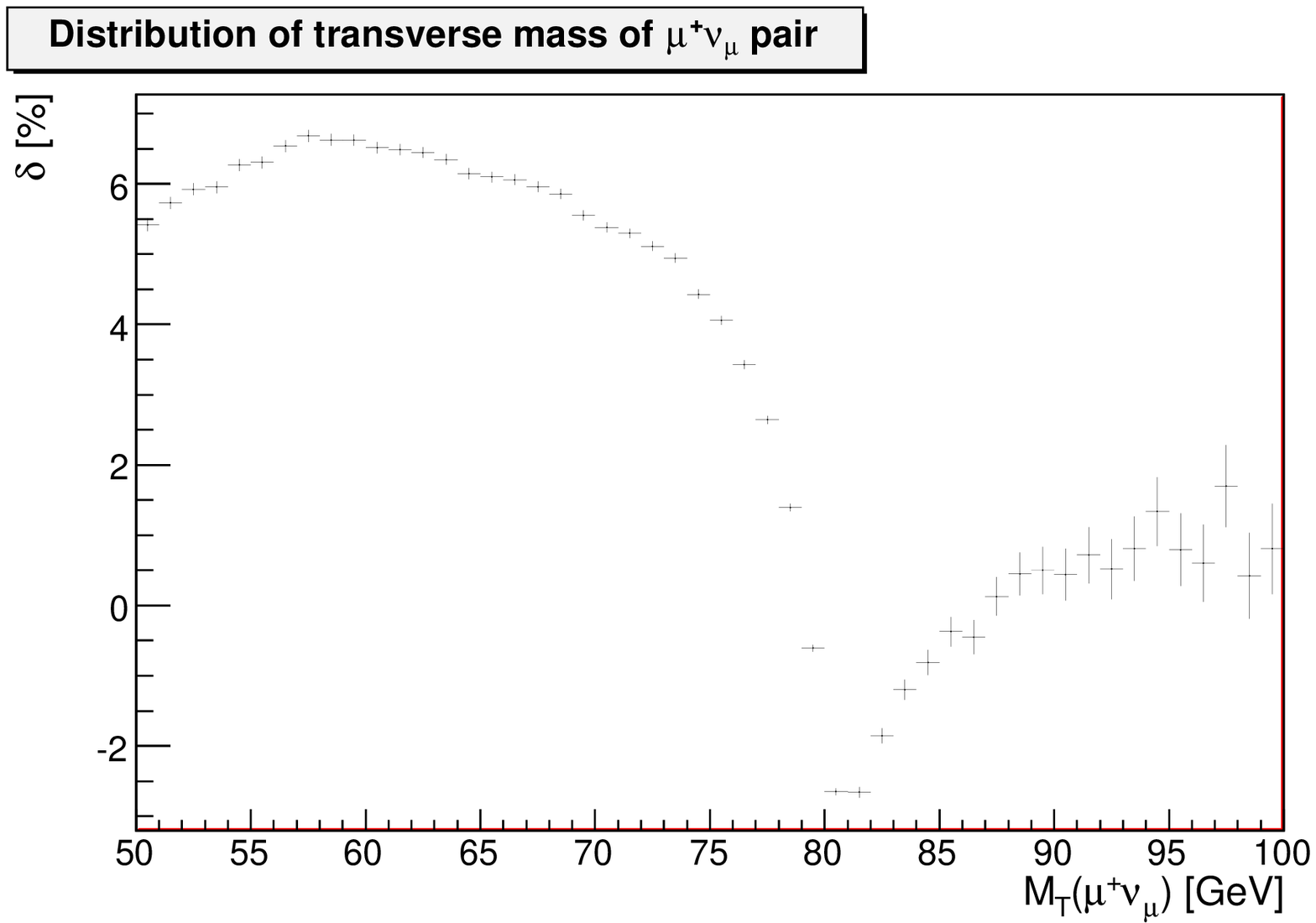}
\caption{The EW NLO distributions of $ M_T^W $ and the relative corrections}
\label{mtw}
\end{figure}

\begin{figure}[!th]
\includegraphics[width=8cm,height=5cm]{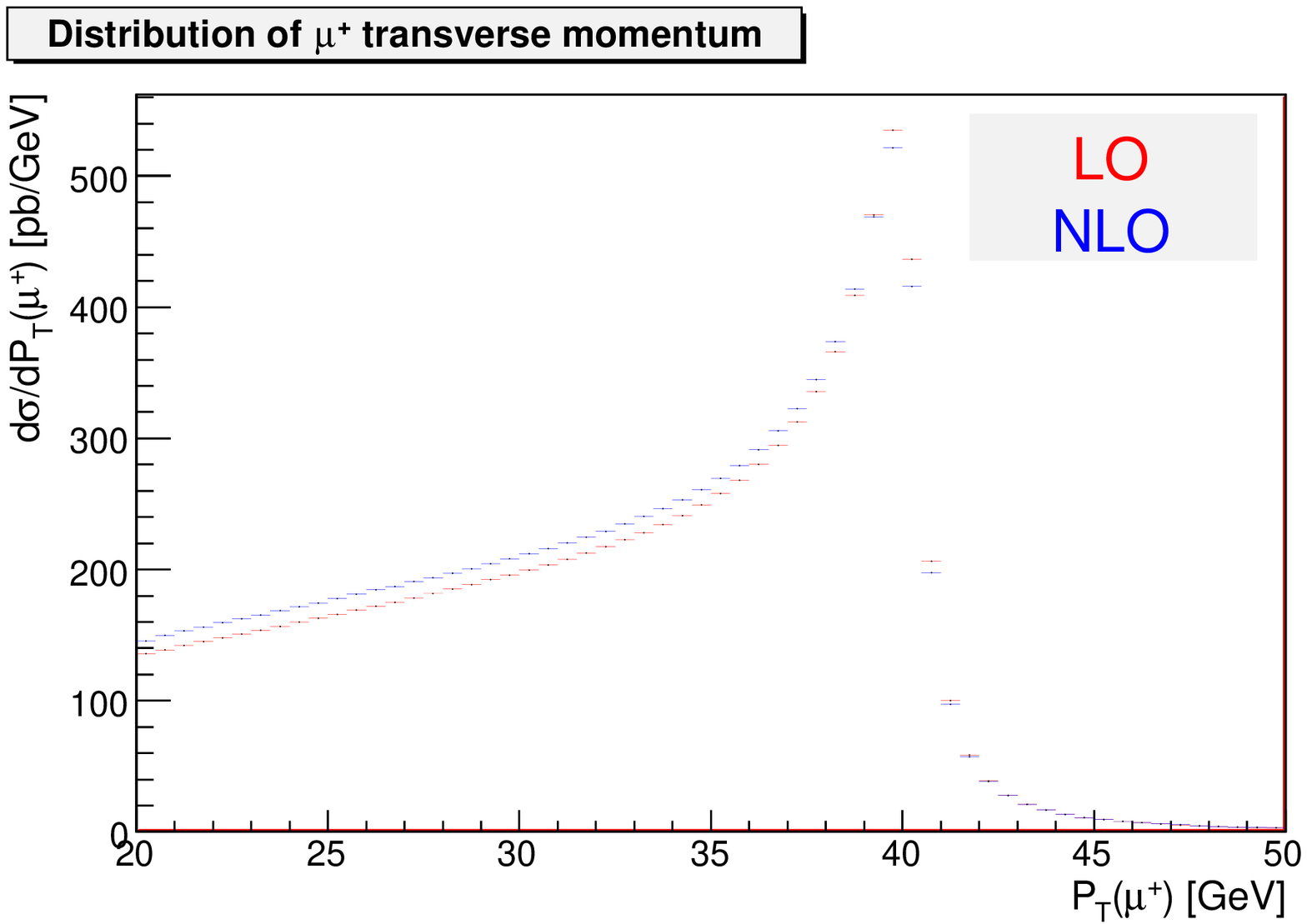}
\includegraphics[width=8cm,height=5cm]{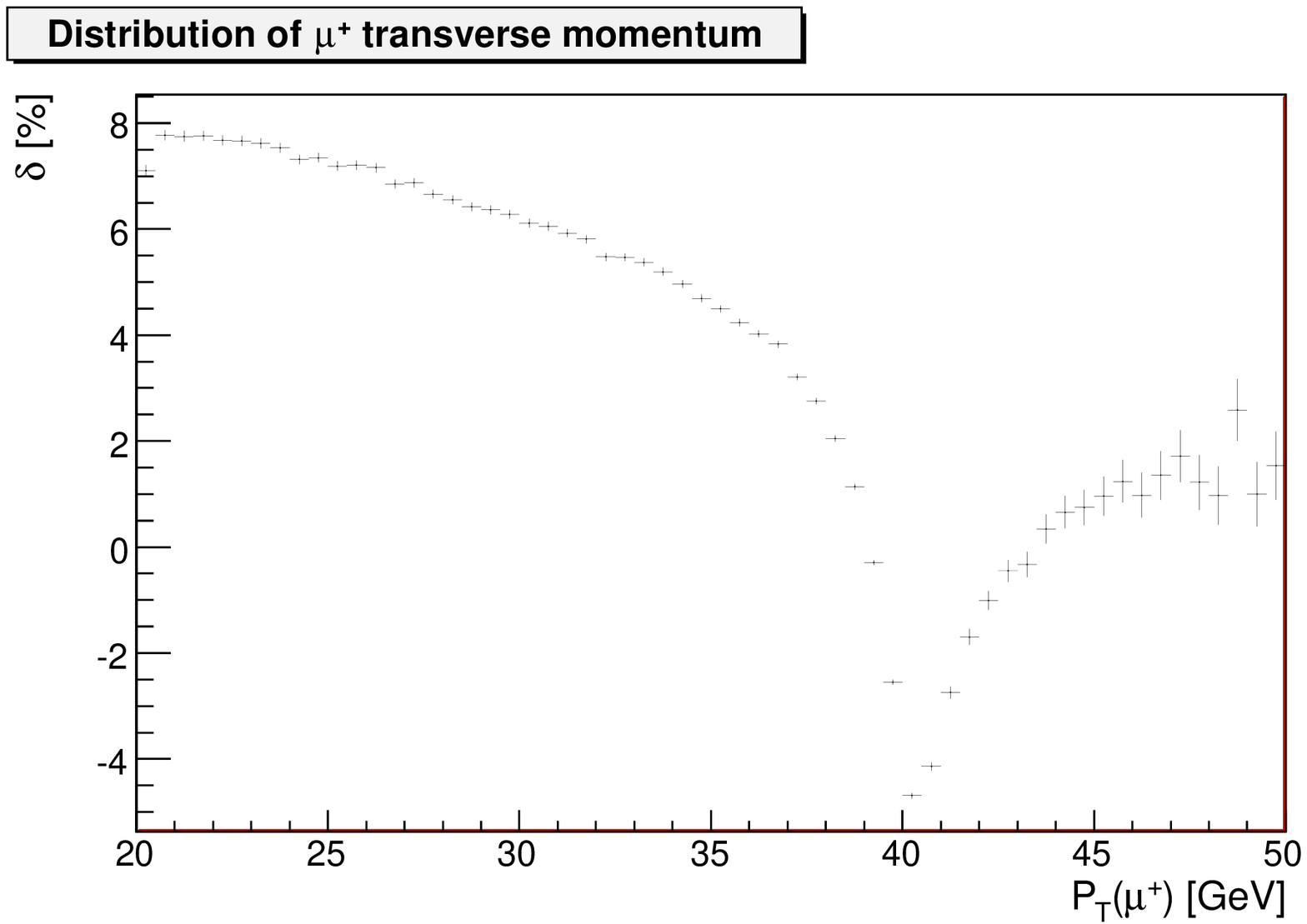}
\caption{The EW NLO distributions of $ p_T^{\mu} $ and the relative corrections}
\label{ptmu}
\end{figure}

Some results obtained with these generators were presented in two talks to ATLAS MC  
group~\cite{Sadykov:2007mc}, \cite{Sadykov:2008mc}. Examples of distributions produced with
help of MC generator for single-W production are shown in Fig.~\ref{mtw} and Fig.~\ref{ptmu}. 

\section{Example of the implementation of the SSFM from {\tt SANC} to {\tt WINHAC}.}
\label{SANCandWINHAG}

The result of the implementation of SSFM for EW RC at one-loop level is presented in
\cite{Bardin:2008fn}.

For the description of {\tt WINHAC} and {\tt SANC} we refer the reader to the literature:
for {\tt WINHAC} to~\cite{Placzek:2003zg}, \cite{WINHAC:MC} and for {\tt SANC} to~\cite{Andonov:2004hi}.
The goals of the work \cite{Bardin:2008fn} were:
\begin{itemize}
\item to check the implementation of SSFM  EW NLO modules
into the framework of {\tt WINHAC} Monte Carlo event generator;

\item to perform a tuned comparison of two codes:\\
1. the standard {\tt SANC} integrator DY\_CC\_VEGAS with a modified treatment of ISR QED corrections;\\
2. the modified {\tt WINHAC}, upgraded with the {\tt SANC} electroweak modules and downgraded to  
the $\cal O(\alpha)$ QED corrections.
\end{itemize}

 We reached the agreement between the {\tt WINHAC} MC event generator and the {\tt SANC} MC integrator 
for the ${\cal O}(\alpha)$ EW corrections to the CC Drell--Yan process at the level of $\sim 0.025\%$.

\begin{figure}[!h]
\begin{center} 
\includegraphics[height=10cm,width=16cm]{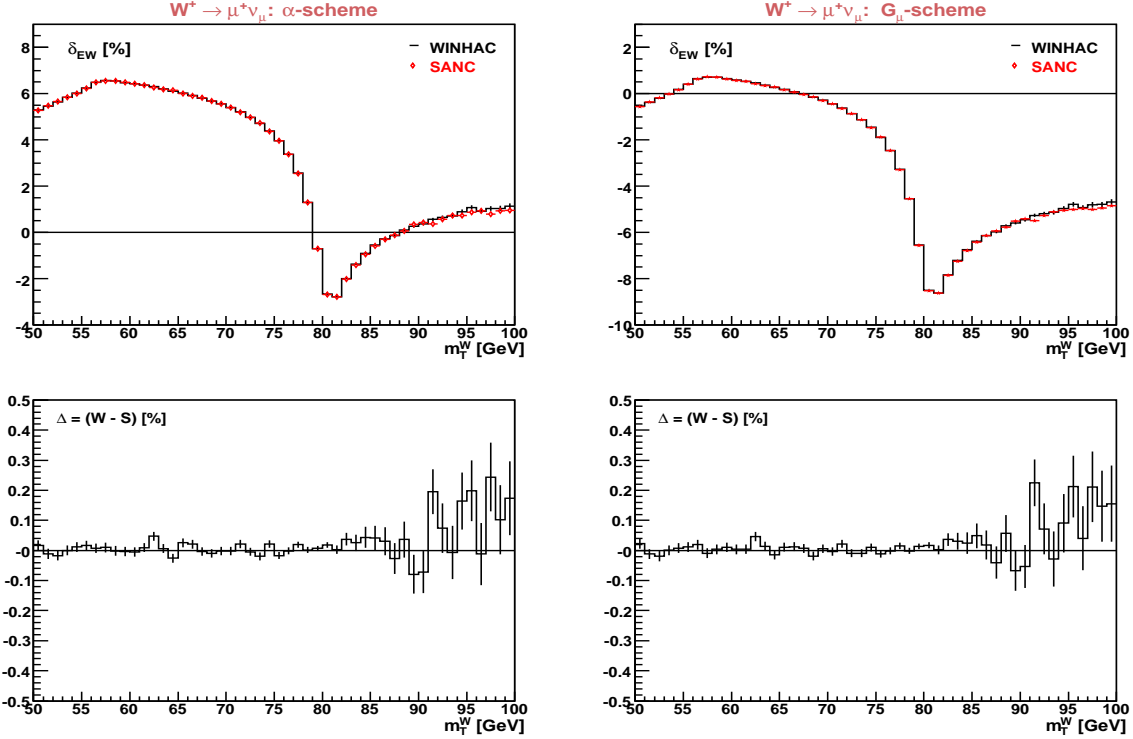}
\end{center}                              
\vspace*{-5mm}
\caption
[The EW NLO distributions of $m^{\rm W}_{\rm T}$ for the muon channel in two schemes and the 
absolute deviations]
{The EW NLO distributions of $m^{\rm W}_{\rm T}$ for the muon channel in two schemes and the 
absolute deviations between {\tt WINHAC} and {\tt SANC}, $\Delta={{\rm W}-{\rm S}}$, both quantities
in [\%].}
\label{fig:mtloop}
\end{figure}

 As an example we demonstrate the comparison of the transverse mass $m^{\rm W}_{\rm T}$
distributions for the muon channel in two popular EW parametrization schemes ($\alpha$ and $G_{\mu}$, 
see e.g. Ref.~\cite{Wackeroth:1996hz}) with ``simplified bare'' cuts, see Fig.~\ref{fig:mtloop}. 
The two upper figures show the RC quantity  {$\delta_{\rm EW}$} in \%,
while the two lower figures show absolute deviations {W$-$S} between the two calculations,
for details see Ref.~\cite{Bardin:2008fn}.


\section{User guide}
 Here we present the technical description of the {\tt SANC} NC and
{\tt SANC} CC packages --- v.1.20, 
intended for calculation of the total  and differential $d\sigma/d\cos\vartheta$ cross sections at 
the partonic level.
Conventions, enumerations and descriptions of the options of the main flags are also given.

The packages can be accessed from project homepages~\cite{SANCsite:2008}.

\subsection{Naming conventions}
In {\tt SANC} we use naming conventions for fields (or particles) shown in
Table \ref{listfilds} where N is the field index (here we present only physical particles, omitting ghost fields)
and in the columns headed ``name'' we show the names used internally in {\tt SANC}. All associated parameter symbols are 
derived from these names. Thus the mass, charge and weak isospin of the electron are denoted {\tt mel}, {\tt qel} and
{\tt i3el}, respectively, also the vector and axial vector coupling constants ({\tt vel, ael}).

\begin{table}[!h]
\caption[List of fields.]
        {List of fields.}
\label{listfilds}
\vspace*{3mm}
\begin{tabular}{||r|c|c|c|c|c|c|c|c|c|c|c|c|c|c||}
\hline
\hline
\multicolumn{3}{||c}{bosons}
&
\multicolumn{9}{|c|}{fermions}
&
\multicolumn{3}{c||}{QCD}\\
\hline    
\multicolumn{3}{||c}{}
&
\multicolumn{3}{|c}{1st generation}
&
\multicolumn{3}{|c}{2nd generation}
&
\multicolumn{3}{|c}{3rd generation}
&
\multicolumn{3}{|c||}{}\\
\hline
   $N$ &  field  &name& $N$ &field&name& $N$ &field &name& $N$&field &name& $N$&field &name \\
\hline
     1 &$A$         & gm  & 11 &$\nu_e$&en&15&$\nu_\mu$ & mn & 19 &$\nu_\tau$&tn& 23 &  g  & gn \\
     2 &$Z$         &  z  & 12 &$e$    & el  & 16 &$\mu$& mo & 20 &$\tau$  & ta &    &     &    \\
$\pm$3 &$W^{\pm}$   &  w  & 13 &$u$    & up  & 17 &$c$  & ch & 21 &  $t$   & tp &    &     &    \\
     4 &$H$         &  h  & 14 &$d$    & dn  & 18 &$s$  & st & 22 &  $b$   & bt &    &     &    \\
\hline
\hline
\end{tabular}
\end{table}

\subsection{Overview of the packages}
As an example, we present an overview of the most recent package {\sf sanc\_cc\_v1.20}~\footnote{The structure 
of the  package {\sf sanc\_nc\_v1.20} is similar.}.
It contains SSFM of two types: the modules {\sf cc\_*\_xy\_zv.f(F)} for 1-loop level EW sub-processes, which
are produced by the {\tt s2n} package of {\tt SANC} project and are governed by {\sf main} file {\sf main\_cc\_1loop\_vegas.F}
written ``by hand''; the modules {\sf cc-qcd\_*\_xy\_zv.f} for 1-loop level QCD sub-processes
and {\sf  cc*\_23\_x\_\_y\_z\_v.f(F)} ({ cc*\_\sf 23\_x\_\_y\_z\_v.f(F)}) for gluon-induced tree level partonic sub-processes
together with corresponding {\sf main} file {\sf main\_cc\_trees\_vegas.F}. The status of QCD SSF(ORM)M is somewhat different 
from EW ones; the exist, however, at present time are not yet put into the system. Nevertheless, SSF(ORTRAN)M were produced from
FORM log-files by mean of the standard {\tt SANC} {s2n} software. 
Next, the package contains declaration files --- {\sf *.h} and libraries --- {\sf *.a}. 
The folder {\sf SancLib\_v1\_02} contains the source files comprising the library for various functions, including Vegas 
integration and the library itself {\sf libSancLib\_v1\_02.a}. 
Here and below ``x,y,z'' and ``v'' stand for the standard {\tt SANC} field indices, i.e.
$12 - e,\;13 - u,\;14 - d, 16 - \mu, 23 - g$, {\it etc}. In Table~\ref{pids} we summarize available in {\sf sanc\_cc\_v1.20}
1-loop level and in Table~\ref{pidstrees} all gluon-induced tree level partonic sub-processes, 
renumbered by the Process IDentifier (PID):

\begin{table}[!h]
\caption[Process IDentifier (PID) and available at 1-loop level CC DY sub-processes.]
        {Process IDentifier (PID) and available at 1-loop level CC DY sub-processes.}
\label{pids}
\begin{center}
\begin{tabular}{||c|c|c||}
\hline
\hline
  PID &     xy\_zv      & Sub-process                 \\
\hline
   1  &   1314\_1112    & $\bar{u} + d \to \bar{\nu}_e + e^-    $ \\
   2  &   1413\_1211    & $\bar{d} + u \to         e^+ + \nu_e  $ \\
   3  &   1314\_1516    & $\bar{u} + d \to \bar{\nu}_{\mu}+\mu^-$ \\
   4  &   1413\_1615    & $\bar{d} + u \to \mu^+  + {\nu}_{\mu} $ \\
   5  &   1314\_1920    & $\bar{u} + d \to \bar{\nu}_{\tau}+\tau^-$ \\
   6  &   1413\_2019    & $\bar{d} + u \to \tau^++ {\nu}_{\tau} $ \\
\hline
\hline
\end{tabular}
\end{center}
\end{table}

 The total set of files inside the package is:\\

\noindent
\underline{Instruction files}:
\begin{verbatim}
README
RELEASE-NOTES
CHANGES 
LICENSE.TXT
INSTALL
FILES 
\end{verbatim}

\noindent
\underline{Declaration files}: 
\begin{verbatim}
s2n_declare.h 
\end{verbatim} 

\clearpage

\begin{table}[!th]
\caption[Process IDentifier (PID) and available at tree level gluon-induced sub-processes.]
        {Process IDentifier (PID) and available at tree level gluon-induced sub-processes.}
\label{pidstrees}
\begin{center}
\begin{tabular}{||c|c|c||}
\hline
\hline
  PID &  23\_x\_\_y\_z\_v & Sub-process                 \\
  PID & (x\_23\_\_y\_z\_v)& Sub-process                 \\
\hline
   7  & 23\_13\_\_14\_11\_12 & $g + u      \to d       + \nu_e + e^+       $\\
   8  & 14\_23\_\_13\_12\_11 & $g + \bar{d}\to \bar{u} + \nu_e + e^+       $\\
   9  & 23\_14\_\_13\_12\_11 & $g + d      \to u       + e^- + \bar{\nu}_e $\\
  10  & 13\_23\_\_14\_11\_12 & $g + \bar{u}\to \bar{d} + e^- + \bar{\nu}_e $\\
  11  & 23\_13\_\_14\_15\_16 & $g + u      \to d       + \nu_{\mu} + \mu^+ $\\
  12  & 14\_23\_\_13\_16\_15 & $g + \bar{d}\to \bar{u} + \nu_{\mu} + \mu^+ $\\
  13  & 23\_14\_\_13\_16\_15 & $g + d      \to u       +\mu^- + \bar{\nu}_{\mu} $\\
  14  & 13\_23\_\_14\_15\_16 & $g + \bar{u}\to \bar{d} +\mu^- + \bar{\nu}_{\mu} $\\
  15  & 23\_13\_\_14\_15\_16 & $g + u      \to d       + \nu_{\tau} + \tau^+ $\\
  16  & 14\_23\_\_13\_16\_15 & $g + \bar{d}\to \bar{u} + \nu_{\tau} + \tau^+ $\\
  17  & 23\_14\_\_13\_16\_15 & $g + d      \to u       +\tau^- + \bar{\nu}_{\tau} $\\
  18  & 13\_23\_\_14\_15\_16 & $g + \bar{u}\to \bar{d} +\tau^- + \bar{\nu}_{\tau} $\\
\hline
\hline
\end{tabular}
\end{center}
\end{table}

\noindent
\underline{Initialization and various input files}: 
\begin{verbatim}
s2n_init.f
sanc_input.h
leshw_input.h
tev4lhcw_input.h
leshw2007_input.h
\end{verbatim} 

\noindent
\underline{SSFM}\hspace{2.5cm} originated from
\begin{verbatim} 					      		   
 cc_ff_xy_zv.F        (FF)
 cc_si_xy_zv.f        (HA)
 cc_[qcd_]br_xy_zv.f  (BR)
[this file contains three SSFM (subroutines) for EW case and two --- for QCD case
 cc_bo_xy_zv(...), cc_br_xy_zv(...), cc_ha_xy_zv_1spr(...)
 cc_qcd_br_xy_zv(...), cc_qcd_ha_xy_zv_1spr(...)]
 cc_[qcd_]ha_xy_zv.f  (MC)
\end{verbatim} 

\noindent
\underline{Main file:} {\sf main\_cc\_1loop\_vegas.F} 
\vspace*{3mm}

As a rule of the {\tt SANC} approach, we subdivide the EW RC into the virtual (loop) ones,
the ones due to soft photon emission, and the ones due to hard photon emission with the aid of
the soft--hard separator $\bar\omega$.
For all {\tt SANC} processes we demonstrate the numerical independence off this auxiliary parameter.
The adopted form of presentation of the differential cross section at the one-loop level in obvious notation is:
\bqa
d\hat\sigma^{\rm 1-loop}=
d\hat\sigma^{\rm Born}+d\hat\sigma^{\rm Subt}+d\hat\sigma^{\rm Virt+Soft}(\bar\omega)+d\hat\sigma^{\rm Hard}(\bar\omega).
\label{xsections}
\eqa
The second term stands for subtraction of collinear quark mass singularities.
It may be $d\hat\sigma^{\rm YFS}$ or $d\hat\sigma^{\msbar\rm{(DIS)}}$, correspondingly 
(see~\cite{Bardin:2008fn} and~\cite{Arbuzov:2005dd}). At the partonic level only $\msbar$ option is realized:
\bqa 
\label{msbarsi}
d\hat{\sigma}^{\msbar} = d\hat{\sigma} - \Delta \hat\sigma^{\msbar},
\eqa
where
\bq
 \Delta \hat\sigma^{\msbar} = \lim_{\bar{\omega}\to 0}
\left[\Delta\hat\sigma^{\rm Virt+Soft}(\bar\omega) + \Delta\hat\sigma^{\rm Hard}(\bar\omega)\right]^{\msbar}.
\label{pSubst}
\eq

\clearpage

It is described in detail in Refs.~\cite{Arbuzov:2005dd},~\cite{Arbuzov:2007db}.
At the hadronic level, where Eq.~(\ref{xsections}) is convoluted with PDF's in the usual way, the other two 
``subtraction'' options (YFS or DIS)  were used.
The terms, made up of the virtual (loop) ones --- $d\sigma^{\rm Virt}$,
the ones due to soft photon emission $d\sigma^{\rm Soft}$ , and the ones due to hard 
photon emission $d\sigma^{\rm Hard}$ are subdivided into ISR, IFI and FSR in accordance with $W$-splitting 
techniques~\cite{Wackeroth:1996hz}.
An auxiliary parameter $\bar\omega$ separates the soft and hard photonic contributions. 

The steps of calculation in the {\sf main\_cc\_1loop\_vegas.F} file are in accordance with Eqs.~(\ref{xsections}):

\begin{itemize}
\item{}step of \underline{declaration} and \underline{initialization}, followed by a {\sf call ProcessInit (pid)},
see Section~\ref{oof} for description of various options of flags;
\item{}step \underline{born} is realized by flag iborn=1,\\
\underline{$d{\hat\sigma^{\rm Born}}$} is computed by integration over $\hat{c}$, see Ref.~\cite{Arbuzov:2007db},
of the \underline{\sf function\_ew\_1c},\\
via {\sf call ProcessBorn (...,born,...)}
        \begin{itemize}
        \item{} and SSFM {\sf call cc$\_$br$\_$xy$\_$zv(...,born,...)}
        \end{itemize}
\vspace*{2mm}

\item{}step \underline{virt+soft} and all subsequent ones are realized by flag iborn=0,\\  
\underline{$d\hat\sigma^{\rm Virt+Soft}$} is computed by integration 
over $\hat{c}$ of the \underline{\sf function\_(ew,qcd)\_1c},\\
via {\sf call ProcessVirt\_(EW,QCD) (...,virt,...)}
        \begin{itemize}
        \item{}and {\tt Virt(Weak)} by SSFM call cc$\_$si$\_$xy$\_$zv(...,sigma) ,\\
              (inside this module there exists a call to SSFM cc$\_$ff$\_$xy$\_$zv(-s,-t,-u))
        \item{}and {\tt Virt+Soft(QED,QCD)} 
             \begin{itemize} 
               \item{} for QED processes by SSFM call cc$\_$br$\_$xy$\_$zv(...,soft,...)
               \item{} for QCD processes by SSFM call cc\_qcd\_br\_xy\_zv(...,soft,...)
             \end{itemize}
        \end{itemize}
\vspace*{2mm}

\item{}step \underline{brdq}, i.e. \underline{$(\Delta\hat\sigma)^{\msbar}$},\\
via {\sf call ProcessBrdq\_(EW,QCD) (...,brdq,...)}\\
a) first part, i.e. \underline{$(\Delta\hat\sigma^{\rm Virt+Soft})^{\msbar}$},
        \begin{itemize}
        \item{} is computed by SSFM call cc$\_$bo$\_$xy$\_$zv(...,born), 
        multiplied by $(\delta^{\rm SV})^{\msbar}$ --- a CC analog of Eq.(19) of~\cite{Arbuzov:2007db}
        \end{itemize}
b) second part, i.e. \underline{$(d\hat\sigma^{\rm Hard})^{\msbar}$} is computed
by integration over $\xi$ of the \underline{\sf function\_(ew,qcd)\_1s}
        \begin{itemize}
        \item{} and by SSFM call cc$\_$bo$\_$xy$\_$zv(...,bornk), 
        using Eq.(12) of~\cite{Arbuzov:2005dd}
        \end{itemize}
\vspace*{2mm}

\item{}step \underline{hard}, i.e. \underline{$d\hat{\sigma}^{\mathrm{Hard}}/\dd \hspr$}
is computed by integration over $\hspr$ of the \underline{\sf function\_(ew,qcd)\_1spr}\\
  via {\sf call ProcessHard\_(EW,QCD)\_1spr (...,hard,...)} 
        \begin{itemize}
        \item{} and SSFM call cc\_[qcd\_]ha\_xy\_zv\_1spr (...,hard)
        \end{itemize}
or alternatively 
\vspace*{2mm}

\item{step \underline{hard}, i.e. \underline{$d\sigma^{\rm Hard}/d\Phi^{(3)}$}}
is computed by integration over 4d-phase space of Eq.~(6)--(7)
        of~\cite{Arbuzov:2007db} of the \underline{\sf function\_(ew,qcd)\_4d}\\                       
 via {\sf call ProcessHard\_(EW,QCD) (...,hard,...)} 
         \begin{itemize}
         \item{} and SSFM call cc\_[qcd\_]ha\_xy\_zv(...,hard)
         \end{itemize}      
\end{itemize}

\noindent
\underline{Main file:} {\sf main\_cc\_trees\_vegas.F} 
\vspace*{1mm}

The presentation of the differential cross section for this case is trivial:
\bqa
d\hat\sigma^{\rm trees}=d\hat\sigma^{\rm Born}.
\label{xsectionstrees}
\eqa
Here the Born-term describes the tree level cross section of one of processes of Table~\ref{pidstrees}.
The subtraction of collinear final quark mass singularities for the time being is applied in integrators and generators 
and is not implemented at the parton level.

\subsection{Flag options\label{oof}}
\begin{description} 
\item[\underline{\bf pid(I)}] --- choice of the 1-loop level sub-process, see Table~\ref{pids}:
\item[\phAV I=1,...,6]
\end{description}

\vspace*{-1mm}
\noindent
or of the tree level ones of Table~\ref{pidstrees}:
\vspace*{-1mm}

\begin{description} 
\item[\underline{\bf pid(I)}] \hspace*{8.5mm} {\bf I=7,...18}
\end{description}

\begin{description} 
\item[\underline{\bf iqed(I)}] --- choice of calculations for QED correction: 
\item[\phAV I=0]  without~ QED correction
\item[\phAV I=1]  with all QED correction
\item[\phAV I=2]  with ISR QED correction
\item[\phAV I=3]  with IFI QED correction
\item[\phAV I=4]  with FSR QED correction
\item[\phAV I=5]  with IFI and FSR QED correction
\end{description}

\begin{description} 
\item[\underline{\bf iew(I)}] --- choice of calculations for EW correction: 
\item[\phAV I=0]  without EW correction
\item[\phAV I=1]  with EW correction
\end{description}

\begin{description} 
\item[\underline{\bf iqcd(I)}] --- choice of calculations for EW correction: 
\item[\phAV I=0]  without QCD correction
\item[\phAV I=1]  with QCD correction
\end{description}

\begin{description} 
\item[\underline{\bf iborn(I)}] --- choice of scheme of calculations of the partonic cross section: 
\item[\phAV I=1]  only Born level
\item[\phAV I=0]  Born + 1-loop virtual corrections
\end{description}

\begin{description} 
\item[\underline{\bf gfscheme(I)}] --- choice of the EW scheme: 
\item[\phAV I=0]  $\alpha$(0)scheme
\item[\phAV I=1]  $G_F $ scheme
\item[\phAV I=2]  $G_F^{'}$ scheme
\end{description}

\begin{description} 
\item[\underline{\bf ilin(I)}] --- choice of the linearization at the calculation of the partonic cross section:
\item[\phAV I=0]  without linearization
\item[\phAV I=1]  {with linearization, {\it i.e.} neglecting spurious terms ${\cal{O}}(\alpha^2)$}
\end{description}

\begin{description} 
\item[\underline{\bf ifgg(I)}] --- choice of calculations of photonic vacuum polarization ${\cal{F}}_{gg}$:
\item[\phAV I=$-$1] {--- $0$}
\item[\phAV I= ~0] {--- $\rm 1$}
\item[\phAV I= ~1] {--- $\rm {1 + k {\cal{F}}_{gg}}$}
\item[\phAV I= ~2] {--- $\rm {1/(1 - k {\cal{F}}_{gg})}$}
\end{description}
with $k=\frac{\ds g^2}{\ds 16\pi^2}$\,.

\begin{description} 
\item[\underline{\bf ihard(I)}] --- types of the hard bremsstrahlung phase-space integrations:
\item[\phAV I=1] {integration over $\hspr$}
\item[\phAV I=4] {4d integration}
\end{description}

\begin{description} 
\item[\underline{\bf isetup(I)}] --- choice of the setup: 
\item[\phAV I=0] {Standard {\tt SANC}}
\item[\phAV I=1] {Les Houches Workshop, 2005}
\item[\phAV I=2] {TeV4LHC Workshop, 2006}
\item[\phAV I=3] {Les Houches Workshop, 2007}
\end{description}


\section{Conclusions}
In this paper we have described the Standard {\tt SANC} FORTRAN Modules and presented examples of their application:
1) they were used in the packages at the parton level for quick studies of different features of the given sub-processes: 
estimates of effects due to variations of input parameters, electroweak schemes, interplay
of different RC contributions (EW-QCD, initial and final state radiation) {\it etc.};
2) they were implemented in the MC generator {\tt WINHAC}, increasing thereby its potential possibilities; 
3) finally, they were used in the {\tt SANC} Monte Carlo integrators and event generators that provide distributions 
of the final state particles with full kinematics, which was interfaced with parton showering codes 
({\tt PYTHIA} and {\tt HERWIG})
and the events can be further processed through the whole experiment simulation environment.

\section*{Acknowledgment}
We are grateful to E.~Uglov for support of the {\tt SANC} Download page.

This work is partly supported by Russian Foundation for Basic Research grant $N^{o}$ 07-02-00932.
One of us (A.A.) thanks the grant of the President RF Scientific Schools 3312.2008.
Two of us (V.K.and R.S) thanks also the grant ``Dinastiya''- 2008.

\bibliographystyle{utphys_spires}
\bibliography{ATLAS_note_DY_bib}
\end{document}